\documentclass[conference]{IEEEtran/IEEEtran}

\usepackage[utf8]{inputenc}
\usepackage{tikz}
\usepackage{amssymb}
\usepackage{setspace}

\graphicspath{{figures/}}

\begin{document}
%
\title{
A Data Fusion System to Study Synchronization in Social Activities
}

\author{\IEEEauthorblockN{Lo\"ic~Sevrin,
        Bertrand~Massot,
		Norbert~Noury,
        Nacer~Abouchi,
        Fabrice~Jumel,
        and~Jacques~Saraydaryan }
\IEEEauthorblockA{Université de Lyon, Lab. INL \& CITI, France\\
Email: loic.sevrin@univ-lyon1.fr}
}

\maketitle


\definecolor{electricblue}{rgb}{0.49, 0.98, 1.0}

\definecolor{SkyBlue}{rgb}{0.53, 0.81, 0.92}

\definecolor{applegreen}{rgb}{0.55, 0.71, 0.0}

\definecolor{c1a}{HTML}{0099CC}
\definecolor{c1b}{HTML}{DBFFDB}
\definecolor{c1c}{HTML}{D4F1FF}
\definecolor{c1d}{HTML}{003399}

\definecolor{blue1}{HTML}{3498db}
\definecolor{blue2}{HTML}{2980b9} 
\definecolor{green2}{HTML}{27ae60}
\definecolor{black1}{HTML}{34495e}
\definecolor{red1}{HTML}{e74c3c}
\definecolor{turquoise2}{HTML}{16a085}

\newcommand{\geturi}[1]{../../figures/#1}
\newcommand{\good}[1]{\includegraphics[width=#1]{\geturi{emoji/good.pdf}}}
\newcommand{\goodI}[0]{\good{0.5cm}}
\newcommand{\sizeI}[0]{0.75cm}
\newcommand{\x}[0]{}
\newcommand{\checkplus}[1]{\includegraphics[width=#1]{\geturi{emoji/check1.pdf}}}
\newcommand{\checkminus}[1]{\includegraphics[width=#1]{\geturi{emoji/check2.pdf}}}


\begin{abstract}

As the world population gets older, the healthcare system must be adapted, among others by providing continuous health monitoring at home and in the city.
The social activities have a significant role in everyone health status.
Hence, this paper proposes a system to perform a data fusion of signals sampled on several subjects during social activities.
This study implies the time synchronization of data coming from several sensors whether these are embedded on people or integrated in the environment.
The data fusion is applied to several experiments including physical, cognitive and rest activities, with social aspects.
The simultaneous and continuous analysis of four subjects cardiac activity and GPS coordinates provides a new way to distinguish different collaborative activities comparing the measurements between the subjects and along time.

\end{abstract}


\IEEEpeerreviewmaketitle

\section{Introduction}
\label{part:intro}

The ageing of the worldwide population is a global issue which should be faced as soon as possible.
The actions which must be taken include the provision of an health monitoring system at home and in the city in order to keep people active as they get older, to reduce the unnecessary placements in nursing homes, the overcosts associated with unforeseen emergency hospitalizations.

The health monitoring should be based on activity monitoring since it represents one visible aspects of the health status. 
A health smart home can include several types of monitoring systems. 
Demiris et Al. described six categories: physiological, functional, safety, security, social and cognitive monitoring \cite{demiris2007technologies}.
In particular, the social aspect is both relevant according to the World Health Organisation \cite{world1950preamble} although sparsely studied, especially considering direct interaction between people. Demiris review does not refer to any study of this kind. 

In order to monitor the social activities of a person, the activities of this person and the ones he interacts with should be monitored, synchronized and merged.
The monitoring systems can be both integrated in the home, and/or embedded on the people.
The main advantage with embedded systems is their ability to be pervasive: 
it allows to follow up the subject anywhere he goes with less risk for missing relevant data.
Several existing systems already combine a variety of sensors, including cardiovascular activity and physical actimetry, to reach a high precision on energy expenditures monitoring on a daily basis, although resulting in bulky systems \cite{7073648,massot:hal-01226432}. 
Another approach is to keep the number of sensors to a minimum in order to only obtain a macroscopic view \cite{6379452}.

This paper presents in part \ref{part:material-methods}, a system and a method to synchronize, centralize and merge heterogeneous data measured on several people along the day using a Body Area Network.
This method is further applied in part \ref{part:experimentation} to study social activities, either physical or cognitive, through the heart rates and GPS positions measurements.

\section{Materials and Methods}
\label{part:material-methods}

\subsection{ROS: a modular data fusion platform}

\begin{figure}[]
\centering
\caption{ROS communication protocol, based on topics}
\label{fig:topics-ros}
\begin{tikzpicture}[xscale=0.5, yscale=0.7]
\tikzstyle{nodeBase}=[draw, rounded corners=3pt];
\tikzstyle{nodeP}=[blue2, nodeBase];
\tikzstyle{nodeS}=[turquoise2, nodeBase];
\tikzstyle{nodeT}=[black, nodeBase];
\tikzstyle{legend}=[midway, below, text width=3cm, align=center];
\node (T) at (0,0) {\includegraphics[width=1.6cm]{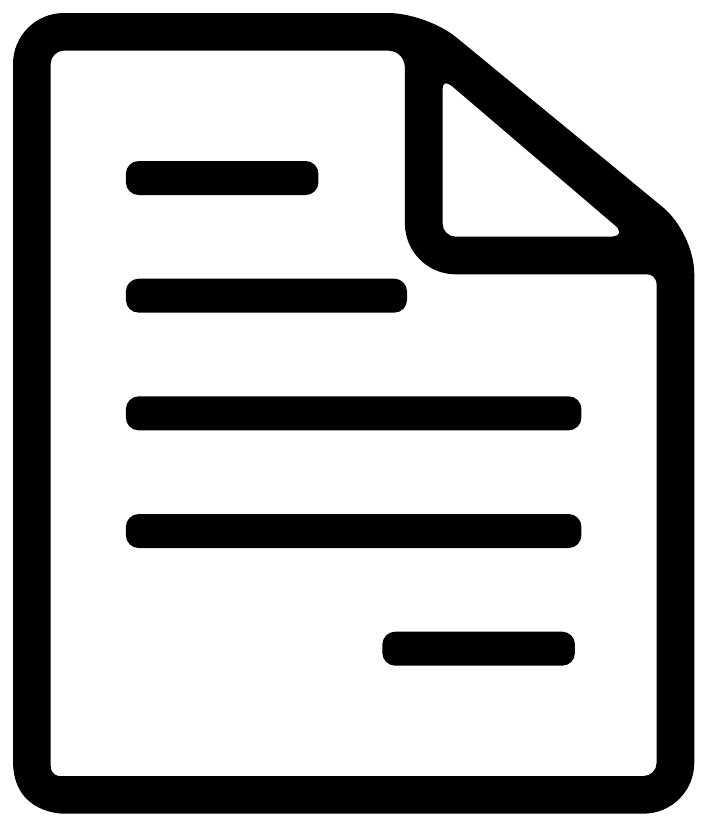}};
\node[nodeS] (S1) at (5,1) [draw,rounded corners=3pt, thick, fill=white]{subscriber node};
\node[nodeS] (S2) at (5,-1) [draw,rounded corners=3pt, thick, fill=white]{subscriber node};
\node[nodeP] (P) at (-5,0) [draw,rounded corners=3pt, thick, fill=white]{publisher node};
\draw[->, thick, nodeS] (T.east) to[bend right=10] (S1);
\draw[->, thick, nodeS] (T.east) to[bend left=10] (S2);
\draw[->, thick, nodeP] (P) to (T.west);

\draw[|-|, nodeP] (-7.5,-2) -- ++(5,0) node[legend]{write data};
\draw[|-|, nodeT] (-2,-2) -- ++(4,0) node[legend]{topic\\(abstraction layer)};
\draw[|-|, nodeS] (2.5,-2) -- ++(5,0) node[legend]{read data};


\end{tikzpicture}
\end{figure}
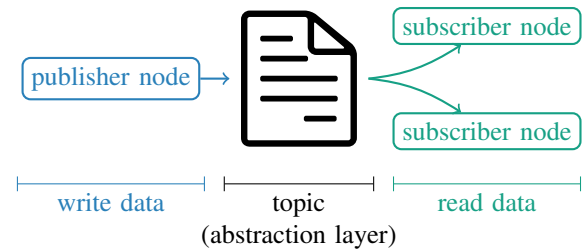

In order to evolve with technology, an activity analysis system must provide a modular design. 
It must be able to handle data from heterogeneous sensors and to communicate over heterogeneous protocols. 
The Robot Operating System (ROS) is a viable solution which provides a high-level communication protocol based on the publisher-subscriber approach.
This software system is based on a cloud of software components called \textit{nodes} running independently on a computer of the network and which can share data by publishing or subscribing to topics. 
The latest are data loggers which can be accessed by any node to read or write shared data.
Hence these topics can be considered as abstraction layers since the publishing node can be changed without any impact on the subscriber side and vice versa (Fig. \ref{fig:topics-ros}).

The livINLab, a living lab at the INL, University of Lyon, has its software architecture based on ROS. This smart environment has been proven to integrate Kinects depth cameras seamlessly, and record the indoor positioning data coming from several depth cameras into a geographic database \cite{Sevrin2015361}.
Thanks to the modular and evolutionary design, the platform can integrate many other sensors and actuators using the tools provided by the developers community and ROS industrial partners (Bosch, Google, etc.), and by developing our customized software \cite{healthcom15submitted}.

In particular, in order to integrate a Body Area Network (BAN), a ROS based platform should be able to communicate over the Bluetooth Low Energy (BLE) protocol which has widely spread in the recent years, driven by the protocol adoption on smartphones. 
In our case, a BLE dongle designed by BlueGiga is used with the provided drivers and libraries. 
The latest can be used from a ROS node, which can publish the incoming data to a topic on the ROS network. 

In short, this ROS node combined with a BLE library provides a bridge between the modular ROS network and any BAN communicating over the BLE protocol. Hence the data coming from many BLE sensors can be centralized in the ROS based system.

\subsection{The REC@MED project}

\begin{figure}
\centering
\caption{ECG electrodes placement on the body}
\label{fig:recamed-sensor}
\includegraphics[width=3cm]{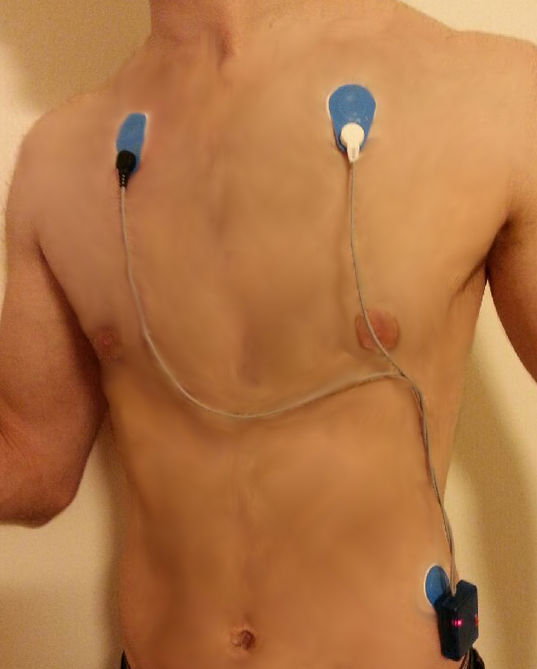}
\end{figure}

The REC@MED project aims to propose a global hardware architecture, with sensors newly developed 
for simultaneous monitoring of cardiac, electrodermal and physical activity, as well as a software platform on smartphone for collecting, storing and passing on data securely \cite{massot:hal-01226432}. 
The first developed sensor measures the electrocardiogram (ECG) of a person using three electrodes placed on the body as on Fig. \ref{fig:recamed-sensor}.

An Android application communicates with the sensor over a BLE connection.
In order to reduce the bandwidth and the power consumption, the complete ECG signal is not sent, but parsed to extract higher level information. 
The computed parameters sent include the R-R intervals, and several parameters of the heart rate variability (HRV) as listed in table \ref{tab:hrv-list}. 
These HRV parameters are computed on five minutes of samples every five minutes. 
The same computation could be performed on a sliding window of five minutes with more processing and electrical power, but considering the sensor lifetime as a key factor, this computation was chosen to be only doable in post processing, using the stored R-R intervals.

\begin{table}
\centering
\caption{Description of short term HRV parameters computed by the REC@MED ECG sensor (source: \cite{massot:hal-01226432})}
\label{tab:hrv-list}
\footnotesize
\begin{tabular}[font size=\tiny]{c|c|c|p{4.5cm}}
\textbf{Variable} & \textbf{Unit} & \textbf{Domain} & \textbf{Description} \\ \hline \hline
SDNN & ms &  Time & Standard deviation of all R-R intervals \\ 
RMSSD & ms & Time & Quadratic mean of differences between successive R-R intervals \\
LF/HF & n.u. & Frequency & Ratio between low-frequency (0.04 to 0.15 Hz) and high-frequency (0.15 to 0.4 Hz) components of the PSD of all R-R intervals \\
LF norm & \% & Frequency & Ratio (expressed as a percentage) between low-frequency components, and the sum of low and high-frequency components of the PSD, i.e. $LF/(LF+HF)$
\end{tabular}
\end{table}

The REC@MED project has been integrated into the ROS network through the bridge previously described.
The  cardiac activity sensor communicates with the phone, wherever the subject goes. 
This way, the system is fully pervasive and monitors the cardiac activity anywhere.
Since the phone includes a GPS receiver, the global position is also measured and recorded.
When the phone can reach the ROS network at the living lab, it starts sending the data it recorded. This way, its database is synchronized with the central one, hosted in the ROS based system.

In order to make this synchronization possible, the throughput must be high enough to transfer a whole day of data within minutes. 
A maximum delay of a few hours could also be acceptable if the transfer id performed during the nighttime.
Considering a transmission at the maximum theoretical rate enabled by the Bluetooth 4.0 (0.27 Mbit/s) and 350 bits of cardiac monitoring data generated every second (for 40 bytes per sample at 66 beats per minute: $350=40\times 8\times 66/60$), the data generated during 24 hours (29 Mbit) would take two minutes to be transfered.
Hence, even if some more data is sent including GPS position or HRV parameters, the sync would take less than 15 minutes to complete.
This result is still acceptable, even if users are more and more familiar with high speed transfers  which could make a dozen of minutes feel like a long time for a sync.

Another solution could achieve a major time synchronization shrink, by using a Wi-Fi connection.
Indeed, Wi-Fi is much faster than Bluetooth 4.0, reaching 1 GBit/s in its 802.11ac release (2014).
But since Wi-Fi and Bluetooth use the same spectrum band (2.4 GHz), many phones use the same chip and antenna for both protocols, which leads to high interferences in many phones.
Since the previously described solution provides a good enough result without this interference problem, it was considered as the best option.

\subsection{Synchronisation and data fusion}

In the REC@MED project, the phone is considered as the master of the BLE network, and also the clock master.
Hence, when the cardiac activity sensor sends a measured R-R interval, it is timestamped at the reception time to avoid the need for a synchronized clock on the sensor. 
This implies that the time between two timestamps does not exactly correspond to the R-R interval itself. 
Hence, the timestamps shall be reconstructed in order to compute the HRV parameters in the post processing step.

When several BAN synchronize their samples with the central database in the ROS based system, the data coming from various BAN must have a common time reference to enable the data fusion processing.
This operation is quite simple since the phones are all connected to the Internet and thus synchronize their clock with a time server. 
With this technique, it remains a delay of up to a couple of seconds between two phones.
But this delay can be neglected when comparing two people cardiac activity or GPS positions since those parameters do not evolve sensibly within a second.

As a conclusion, several people, each being equipped with the REC@MED BAN, can now be monitored together using the centralized database. 
This data fusion is enabled by both the time synchronization of all BAN and by the bridge provided between any Bluetooth device (phones in our case) and the ROS based system and its central database.
The data fusion of cardiac activity and positioning of several people will be exploited in the following experiments social involving social activities.

\section{Experimentation and Validation}
\label{part:experimentation}

\subsection{Fusion of the GPS positions of several people}

The interactions between people mostly occur when they are located in a common area (spatial and temporal unity). 
Therefore, a first data fusion can be operated using the GPS positions of several subjects (four in our case).
These coordinates are recorded by the users phones which include a GPS receiver.
More precisely, the coordinates provided by an Android phone come not only from the GPS chip, but also from the reachable Wi-Fi and GSM networks which are identified by Google.

Given the variety of coordinates sources, and the diversity of environments (mainly indoor or outdoor), the precision estimated by the positioning system varies a lot, from 4 m to 1200 m in our experimental data.
Hence, a collaboration between people following the same trajectory is hard to track. 
Nonetheless, a common unit of space and time can be extracted from these positions. 
In our experiment, four people go from the office to a restaurant 200 m away, and come back after lunch. The resulting map shows all points are in the same area, even if the trajectory cannot be correctly inferred. 

\begin{figure}
\centering
\caption{People positions merged on a unique map showing a unity of space over time, but no precise trajectory between the office and the restaurant (each color represents a different subject)}
\label{fig:gps-fusion}
\begin{tikzpicture}
\def\li{8};
\tikzstyle{annonate}=[black1, fill, opacity=0.9];
\node at (0,0) {
\includegraphics[width=\li cm]{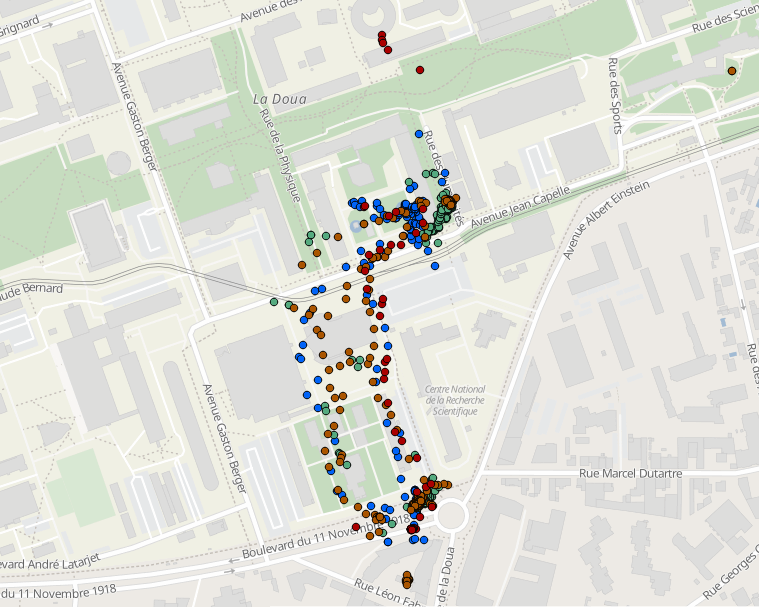}};
\begin{scope}[scale=\li / 20]
\draw[annonate] (1,3) circle (0.5) ++(1,0) node[anchor=west] {office};
\draw[annonate] (-1,-1) circle (0.5) ++(1,0) node[anchor=west] {restaurant};
\end{scope}
\end{tikzpicture}
\end{figure}

\subsection{Fusion of the GPS positions and cardiac activity}

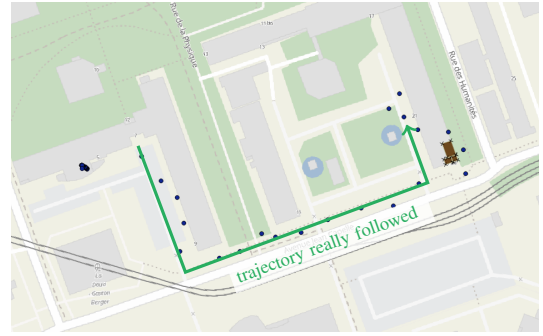
\begin{figure}
\centering
\caption{Recorded coordinates on the return path from the building on the left to the office on the right (blue points), and the real trajectory (green path)}
\label{fig:trajectoire-bertrand-bp}
\begin{tikzpicture}
\def\sca{7};
\node at (0,0) { \includegraphics[width=\sca cm]{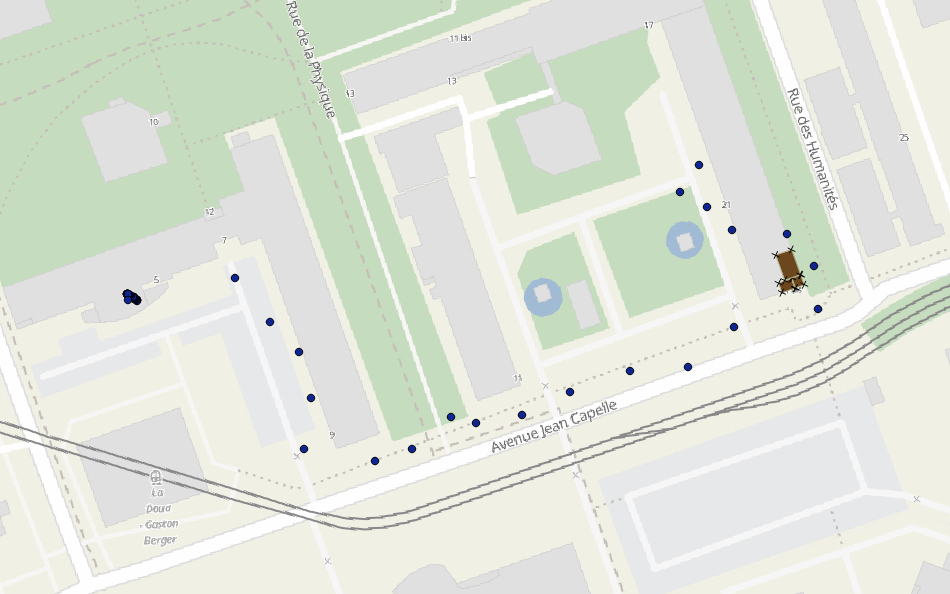}} ;
\begin{scope}[scale=0.1*\sca]
\draw[->, very thick, green2] (-2.6, 0.4) -- (-1.7,-2) -- (2.9, -0.3) -- (2.5, 0.8);
\draw[rotate=20, white, fill, opacity=0.8] (1, -1.5) ++(-2.5,-0.5) rectangle ++(4,0.5); 
\node[green2, rotate=20] at (1, -1.5) {{\scriptsize trajectory really followed}}; 
\end{scope}
\end{tikzpicture}
\end{figure}

The time synchronization between coordinates and cardiac activity can also be used to operate data fusion between a person trajectory and the variation of his heart rate.
This second experiment proposes to study simultaneously the cardiac activity and movement for one subject when the GPS signal is exploitable.
In that scenario, a subject leaves the office, walks to another building where he changes levels by climbing the stairs. The return path of the trajectory, according to the GPS recording is displayed in Fig. \ref{fig:trajectoire-bertrand-bp}.
In this scenario, the measured coordinates are very close to the real ones when the subject is outside.

\begin{figure}
\centering
\caption{The subject heart rate variability (SDNN) drops when the subject moves from one building to another, especially when climbing the stairs}
\label{fig:sdnn-bertrand}
\begin{tikzpicture}
\def\sca{8};
\node at (0,0) {\includegraphics[width=\sca cm]{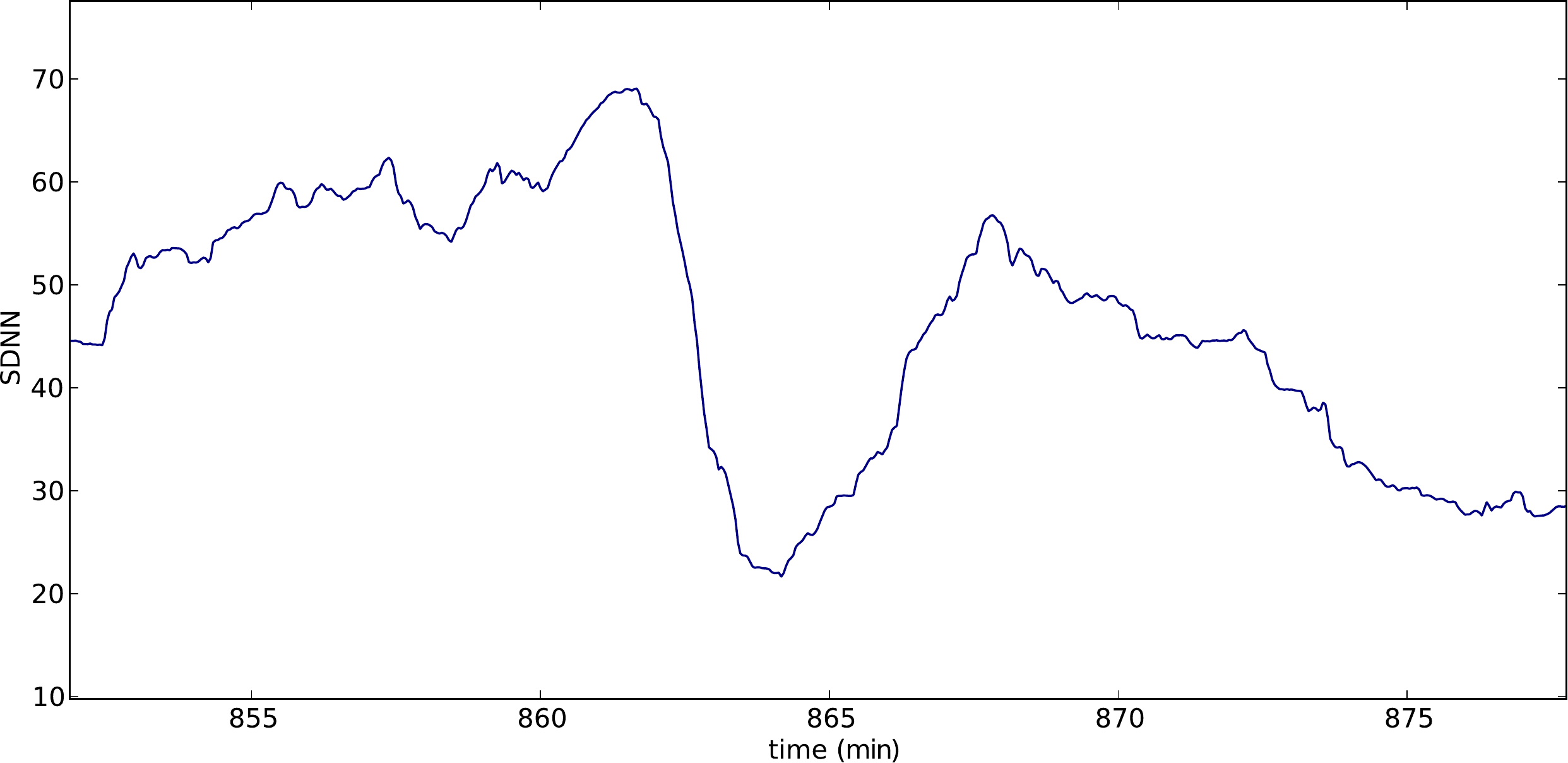}} ;
\begin{scope}[scale=0.1*\sca]
\draw[thick, green2] (-1,0) rectangle (1,-1.5);
\node[green2, font=\scriptsize, align=center] at (-2.7, -0.75) 
{HRV drops with\\intense physical effort}; 
\end{scope}
\end{tikzpicture}
\end{figure}

Once the movement is detected, the heart rate variability of the same time sequence can be drawn as on figure \ref{fig:sdnn-bertrand}.
The drop in HRV is clearly visible on the chart when the subject increases his physical effort (climbing the stairs). 
The fusion of both the trajectories and the cardiac activity can be a good help to classify physical efforts and avoid mixing them with cognitive efforts described in the next section.

\subsection{Fusion of the cardiac activities of several people}

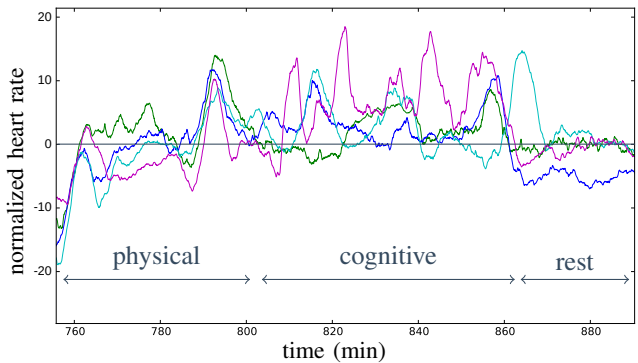
\begin{figure}
\centering
\caption{The normalize HR of the four people over various tasks, social or not. The three activities can be identified only with the fusion of several users data. The black horizontal line is the median of all signals.}
\label{fig:hr-all}
\begin{tikzpicture}
\def\sca{8};
\node at (0,0) {\includegraphics[width=\sca cm]{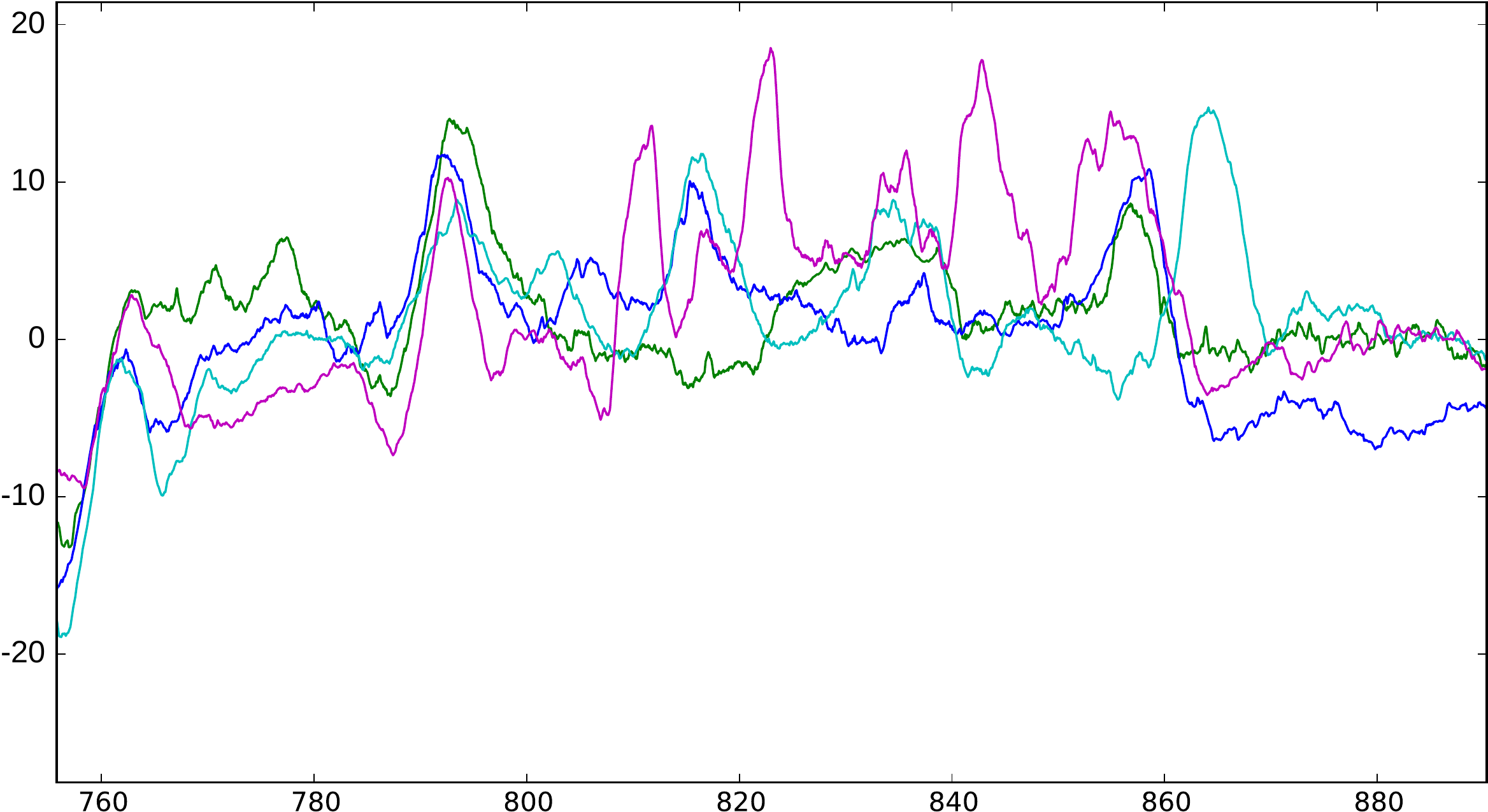}} ;
\begin{scope}[scale=0.1*\sca]
\node[rotate=90] at (-5.3,0) {\small normalized heart rate}; 
\node[] at (0,-3) {\small time (min)}; 
\draw[black1] (-4.6,0.45) -- ++(9.6,0);

\tikzstyle{legend}=[black1,<->];
\draw[legend] (-4.5,-1.8) -- ++(3.1,0) node[midway, above] {physical};
\draw[legend] (-1.2,-1.8) -- ++(4.2,0) node[midway, above] {cognitive};
\draw[legend] (3.1,-1.8) -- ++(1.8,0) node[midway, above] {rest};
\end{scope}
\end{tikzpicture}
\end{figure}

The last part of the performed experiments was designed to study the impact of cognitive social activity on the cardiac activity, and to compare it to the impacts of physical activity and rest.
The cardiac activities of four people are recorded with a common time reference as explained in part \ref{part:material-methods}.
The group of subjects go together at the restaurant. Then they come back and play cards (a strategic French game called Coinche \cite{website:belote}), and finally go back to work.
Hence this scenario is composed of three activities: 
\begin{itemize}
\item a social physical activity (walking)
\item a social cognitive activity (playing cards)
\item a rest period (working alone at a desk)
\end{itemize}
When the heart rates of the four subjects are drawn (Fig. \ref{fig:hr-all}) the three steps can be identified. The HR signals represented are averaged over a sliding window of five minutes and lowered by their median value for easier comparison.

Several observations can be made from this experiment.
Firstly, the rest period can be described as quiet compared to the two others. 
No peak of HR can be seen for the four subjects which are working quietly sitting at their desk.
Secondly, the HR rises quickly, and simultaneously for each subject when they perform a similar physical activity. 
There are two peaks corresponding to the two journeys.
Thirdly, the cognitive collaborative activity causes a higher hear rate than at rest, comparable with the physical activity impact.
Lastly, while playing cards, the people react differently to the same amount of stress induced by the brain activity, handling it an individual way.
Also, with the chosen game, the pressure can be on only one or two players who have strategic cards, which increase the inter-subject variability.
Hence, this last point opens a way to separate a social cognitive activities (playing cards) and a social physical activity (walking together).

\section{Discussion and Conclusion}
\label{part:conclusion}

This paper proposes an easy way to synchronize and centralize heterogeneous data into a database thanks to a ROS based system.
Data fusion can then be performed on the centralized measurements in order to study the subjects cardiac activity and location simultaneously and not only for one single person at a time.
The merged data enabled collaborative activity analysis. 

In the selected scenarios, a physical social effort and a cognitive one can have a similar impact on the cardiac activity. 
In order to differentiate the two, several approach were proposed. 

Firstly the person trajectory was used through the phone positioning system to detect whether the subject was in movement.
This approach has shown good results for outdoor positioning when the GPS chip can provide precise coordinates (resolution under 20 m), but this ideal case is quite unusual. 
Hence inertial sensors or indoor positioning systems (available at the living lab) could provide a more precise analysis in further experiments.
The extensible and modular architecture used to operate the data fusion can easily integrate new sensors. Hence this evolution to a more complete system which could overcome the current limitations will be fluent.

Secondly, the comparison of the cardiac activities of several people performing the same collaborative activity showed a high inter-subject variability for a cognitive social task, whereas the same people behaved (from the cardiac point of view) in a similar manner when walking (social physical activity).
Especially the number an intensity of the HR peaks are very subject dependent. 

At a glance, the use of synchronized data measured during collaborative activity highlighted new comparison parameters that were not visible when studying a single subject alone.
This information could also be used to learn how a person behaves during a social activity, compared to others. 
This collaborative learning could lead to a better understanding of a subject normal behavior to better detect a potential shift in the daily activity which is an important aspect of health monitoring. 
Nonetheless, this results are still preliminary and should be extended before planning such a monitoring.

\bibliographystyle{IEEEtran}
\bibliography{../../biblio/ref2}

\end{document}